\documentclass[conference]{IEEEtran}
\IEEEoverridecommandlockouts
\usepackage{amsmath,amssymb,amsfonts}
\usepackage{algorithmic}
\usepackage{graphicx}
\usepackage{textcomp}
\usepackage{xcolor}
\usepackage{times}
\usepackage{epsfig}
\usepackage{kotex}
\usepackage{booktabs}
\usepackage[normalem]{ulem}
\usepackage{tikz}
\usepackage{cite}
\usepackage{caption}
\usepackage{subcaption}
\usepackage{array}
\usepackage{multirow}
\usepackage{hyphenat}
\usepackage{balance}

\def\BibTeX{{\rm B\kern-.05em{\sc i\kern-.025em b}\kern-.08em
    T\kern-.1667em\lower.7ex\hbox{E}\kern-.125emX}}

\begin{document}

\title{Active Authentication via Korean Keystrokes Under Varying LLM Assistance and Cognitive Contexts \thanks{This paper has been accepted for publication at IEEE ICMLA 2025. The code and dataset are available via the following GitHub repository.\\ https://github.com/rajeshjnu2006/Korean-keystrokes-auth-icmla2025}}


\author{Dong Hyun Roh \\
Bucknell University, USA\\
{\tt\small dhr014@bucknell.edu}
\and
Rajesh Kumar\\
Bucknell University, USA\\
{\tt\small rajesh.kumar@bucknell.edu}
}

\maketitle
 
\begin{abstract}

Keystroke dynamics is a promising modality for active user authentication, but its effectiveness under varying LLM-assisted typing and cognitive conditions remains understudied. Using data from $50$ users and cognitive labels from Bloom's Taxonomy, we evaluate keystroke-based authentication in Korean across three realistic typing scenarios: bona fide composition, LLM content paraphrasing, and transcription. Our pipeline incorporates continuity-aware segmentation, feature extraction, and classification via SVM, MLP, and XGB. Results show that the system maintains reliable performance across varying LLM usages and cognitive contexts, with Equal Error Rates ranging from 5.1\% to 10.4\%. These findings demonstrate the feasibility of behavioral authentication under modern writing conditions and offer insights into designing more context-resilient models. 
\end{abstract}

\begin{IEEEkeywords}

Active authentication, keystroke dynamics, behavioral biometrics, LLM-assisted typing, cognitive load, Korean language, user verification, context-aware security. 
 
\end{IEEEkeywords}
\section{Introduction}

Reliable user authentication is critical for ensuring user privacy and data security \cite{DARPASecurity}. In recent years, authentication systems based on behavioral biometrics have gained significant attention for their ability to enable continuous, passive user authentication \cite{BehavioralBiometricsSurvey}. Commonly studied behavioral traits include typing, swiping, walking, voice, and general body motion. In particular, typing patterns are analyzed using keystroke dynamics, which measures the timing between key presses and releases, capturing \textit{how} a user types rather than \textit{what} the user types \cite{shadman2023, teh2013, kumar2016}. 

Keystroke dynamics is widely valued for being non-intrusive and cost-effective, as it requires no additional hardware beyond a standard keyboard. Furthermore, previous studies have consistently found that imitating another person's typing behavior is extremely difficult, making keystroke dynamics suitable for a wide range of applications \cite{HassanMimicry2018,HassanMimicry2020,plurilockKeystrokeDynamicsDiffMimic}. For example, it has been used for active user authentication and identification \cite{joyce1990, leggett1988, fabianmonrose2000, plurilockKeystrokeDynamicsDiffMimic, HassanMimicry2018, HassanMimicry2020, KeystrokeVideo, KeystrokeSound}, authorship attribution, fake profile detection~\cite{kuruvilla2024}, and inference of soft biometrics \cite{udandarao2020}.

However, typing behavior may vary across languages due to differences in scripts, key combinations, and the use of modifier keys. This variation highlights the need to extend keystroke dynamics research to a broader range of languages, as most existing work remains focused on English. Although several keystroke-based studies have explored non-English contexts—including Arabic~\cite{ArabicKeystrokes}, Chinese~\cite{Multi-Keyboard-Bilingual}, French~\cite{French2017}, Italian~\cite{ItalianKeystrokes2011}, Japanese~\cite{Japanese2009}, Korean~\cite{Junhong2020}, Romanian~\cite{RomanianKeystrokes}, and Russian~\cite{RussianKeystrokes}—the overall body of research remains limited, and many languages are yet to be studied.

Beyond language coverage, important contextual factors remain underexplored in keystroke dynamics research. One such factor is the growing use of Large Language Models (LLMs), such as ChatGPT, which are reshaping how people write and communicate. Users are increasingly adopting LLM-generated content through paraphrasing or transcription, which could potentially alter their natural typing patterns. This shift presents an opportunity to examine keystroke-based authentication in LLM-mediated writing scenarios—an area that has not yet been explored in prior research. Another underexplored dimension is the impact of cognitive load on typing behavior in the context of user authentication, despite evidence that variation in cognitive demand can affect typing pattern~\cite{sellstone2023, balagani2013, vrij2006, zhang2021, hayes2012, flower1981}.

To this end, this paper makes the following contributions:

\begin{enumerate}
\item Investigates keystroke-based authentication performance across three realistic typing scenarios: bona fide composition, LLM-response paraphrasing, and LLM-response transcription.
\item Examines the influence of cognitive load, modeled using Bloom's Taxonomy (remember, understand, apply, analyze, evaluate, create), on authentication performance.
\item Conducts ablation studies on window length, overlap, feature selection, scenario-aware, scenario-unaware, cognition-aware, cognition-unaware, and cross-cognition settings with three widely studied machine learning classifiers (SVM, MLP, and XGB) in the context of active user authentication.   
\end{enumerate}

\begin{table*}[ht]
\centering
\scriptsize
\caption{Comparison of Keystroke Dynamics-based Authentication Studies in the Korean Language.}
\label{tab:kda-comparison}
\renewcommand{\arraystretch}{1.5}
\begin{tabular}{p{2cm} p{4.2cm} p{4.2cm} p{4.2cm}}
\toprule
\textbf{Characteristic} & \textbf{Proposed Work} & \textbf{Mobile FACT \cite{Junhong2020}} & \textbf{Free-text Desktop KDA\cite{Junhong2018Kdua}} \\
\midrule
Goal &
Evaluate the effects of the typing scenario and cognitive load on KDA &
Assess KDA performance on smartphones&
Evaluate KDA performance in unconstrained, free-text conditions \\
Device &
Laptop &
Mobile &
Desktop \\
Language &
Korean only &
Korean and English &
Korean and English \\
Classifiers &
SVM, XGB, MLP &
KS Test, Cramér–von Mises criterion &
One-Class SVM, k-NN, K-Means\\
\uline{Scenarios} &
Bona fide writing, paraphrasing LLM responses, transcribing LLM output &
User-specific script transcription &
User-specific script transcription \\
\uline{Cognition info.} &
Modeled using Bloom's Taxonomy &
Not considered &
Not considered \\
\bottomrule
\end{tabular}
\end{table*}

\section{Related Work}
 
Keystroke dynamics has long been explored as a behavioral biometric for both login-time and active or continuous authentication. Early studies relied on fixed-text input and simple timing features such as key hold and digraph latency~\cite{joyce1990, leggett1988}, while more recent work has focused on free-text input, machine learning classifiers, and adaptive feature selection~\cite{teh2013, kumar2016, fabianmonrose2000, plurilockKeystrokeDynamicsDiffMimic}. Despite promising performance, most systems assume stable input conditions and overlook the variability introduced by cognitive load or LLM integration in the writing process.

While a growing number of studies have attempted to extend keystroke authentication to languages beyond English \cite{ArabicKeystrokes, Multi-Keyboard-Bilingual, French2017, Japanese2009, RomanianKeystrokes, Junhong2020, ItalianKeystrokes2011, RussianKeystrokes}, the majority rely on fixed-text input and rarely address cognitive context variation and LLM integration.

In the context of Korean keystroke dynamics, Kim et al. ~\cite{Junhong2020} introduced FACT, an authentication framework for mobile devices that incorporates heterogeneous features—including timing, touch coordinates, and motion sensor data—from free-text input. The system achieves low equal error rates (below 1\%) and supports both login-time and post-login user verification without requiring additional hardware. Additionally, Kim et al.~\cite{Junhong2018Kdua} proposed a user-adaptive feature extraction method based on relative digraph timing, combined with novelty detection models. Their approach significantly outperformed fixed-feature baselines, highlighting the value of personalized behavioral features in free-text scenarios.
These works, however, do not consider LLM-related writing behaviors or cognitive diversity. A comparative summary of our system and these prior efforts is presented in Table~\ref{tab:kda-comparison}.

Typing behavior is known to vary in relation to task complexity and cognitive state. Prior research has shown that higher cognitive load can alter keystroke rhythm, pause duration, and error rates~\cite{vrij2006, flower1981, balagani2013, hayes2012}. Despite this, most authentication systems treat typing behavior as context-invariant, limiting robustness in real-world use cases.  

The recent availability of LLMs has further complicated input dynamics. Users now frequently paraphrase or transcribe LLM-generated text, which affects both writing intent and motor execution. Roh et al.~\cite{roh2025llm} introduced a Korean-language keystroke dataset capturing this shift. Their study focuses on classifying responses as bona fide, LLM-response paraphrasing, and transcribing, under cognitive conditions derived from Bloom's Taxonomy. However, the dataset has not been analyzed within the authentication context. 

Another closely related work includes studies by Kundu et al.~\cite{kundu2024} and Crossley et al.~\cite{crossley2024}, who explore keystroke patterns in the context of LLM-generated writing. Kundu et al.~\cite{kundu2024} train stylometric and timing-based classifiers to detect assisted writing but observe substantial degradation across users and tasks. Crossley et al.~\cite{crossley2024} focus on timing patterns in transcription, showing how linear typing diverges from cognitively engaged composition. Both works highlight behavioral shifts resulting from the use of LLMs, but do not address authentication or cross-context modeling.

In contrast, our work investigates active user authentication under LLM assistance and cognitive variation using Korean free-text (bona fide), fixed-text (LLM-response transcription), and a mix of fixed and free-text (LLM-response paraphrasing) typing conditions. To our knowledge, this is the first study to explore user authentication under finer-grained LLM mediation, distinguishing between transcription and paraphrasing, while considering cognitive-load impact. 

\section{Methodology}
\subsection{The authentication pipeline} 
\begin{figure}[htp]
    \centering
    \includegraphics[width=3.56in, height=2.1in]{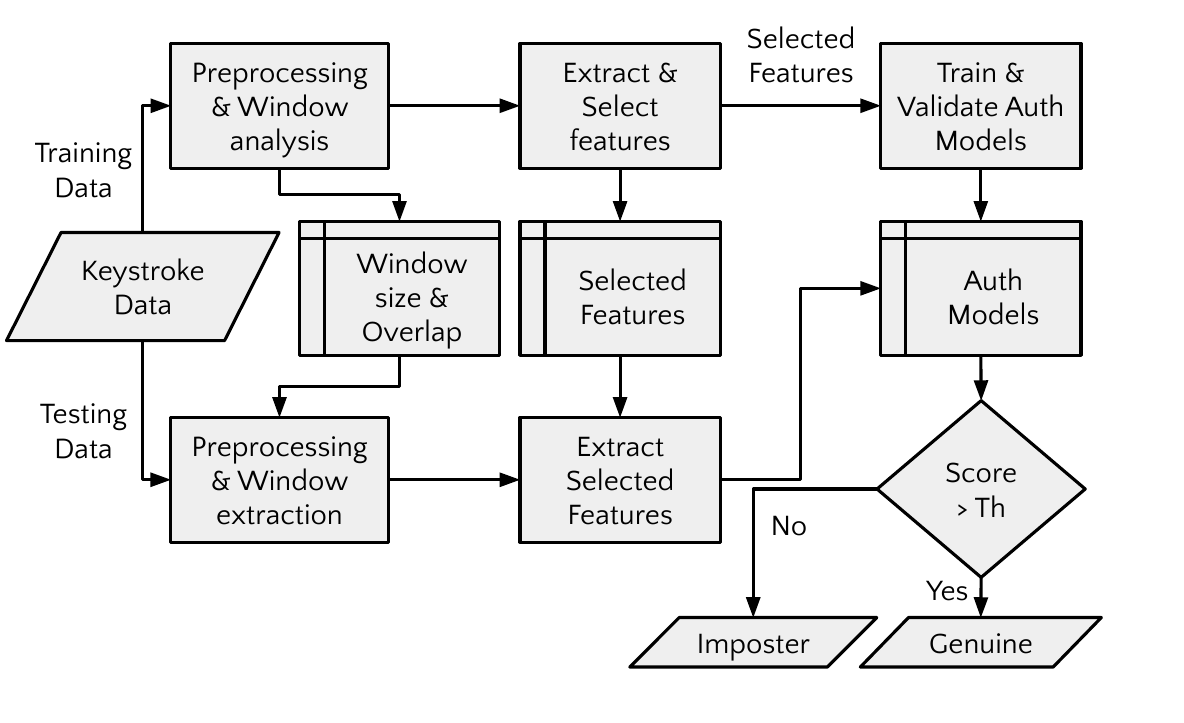}
    \caption{The keystroke-based authentication pipeline begins by structuring raw training and testing data into key event sequences and identifying temporal discontinuities. It then segments each sequence using a sliding window approach and extracts Key Hold Time (KHT) and continuity-aware Key Interval Time (KIT) features. In the enrollment phase, the pipeline applies Minimum Redundancy Maximum Relevance (MRMR) to select informative features from training data and trains the authentication models. During the authentication phase, it extracts the same features from test data. It uses the trained model to classify the user as a genuine or an impostor based on a decision threshold. We use this pipeline for evaluating the authentication model for each scenario and cognitive load condition.}
    \label{fig:pipeline}
\end{figure}

Figure \ref{fig:pipeline} illustrates the authentication pipeline for the proposed keystroke-based user authentication system. The pipeline is designed to support all evaluation settings in this study, including scenario-aware, scenario-unaware, cognition-aware, and cognition-unaware configurations. It processes raw keystroke sequences, accounts for temporal discontinuities, and segments input using a sliding window mechanism to support active authentication. It also incorporates feature selection to enable flexible and robust model training and evaluation. The pipeline consists of five core components: (1) preprocessing and windowing, (2) feature extraction, (3) feature selection, (4) model training and validation, and (5) classification and evaluation.

\subsection{Dataset}

We use the dataset recently proposed by Roh et al. \cite{roh2025llm} as it aligns closely with the objectives of our study. First, the dataset comprises keystroke data in the Korean language, allowing us to expand the language scope of existing keystroke-based authentication studies, which have remained largely confined to English. Furthermore, although a few Korean keystroke datasets already exist, they primarily capture transcription of predefined texts or participant-specific scripted input, not accounting for LLM-mediated writing scenarios. In contrast, the selected dataset captures keystroke data across three distinct writing scenarios—bona fide writing, ChatGPT-response paraphrasing, and ChatGPT-response transcription—better reflecting common typing behaviors associated with the use of LLM and allowing a more nuanced authentication analysis. Lastly, the selected dataset incorporates a cognitive load dimension based on Bloom's Taxonomy, capturing keystroke data for each cognitive process. This supports both cognition-aware and cognition-unaware modeling of typing behavior in the context of user authentication. 

The dataset comprises keystroke data from 69 users. For our study, we focus on a subset of 50 users for whom both Phase 1 and Phase 2 data are available. By using Phase 1 data for training and Phase 2 data for testing our authentication models, we mitigate data leakage risks and enhance the generalizability of our authentication model.

Each Phase consists of three typing scenarios: (1) bona fide typing, (2) ChatGPT-response paraphrasing, and (3) ChatGPT-response transcribing. Within each scenario, participants responded to six questions, each corresponding to a particular cognitive level as defined by Bloom's Taxonomy. Additionally, Phases 1 and 2 used different question sets to minimize memorization effects, while keeping the required word count and data collection environment identical. 

All six questions were presented together in each typing scenario, which led some participants to revisit and revise previously answered questions within the same typing scenario instead of completing the questions in a strict linear order. This resulted in temporally fragmented keystroke event sequences for individual questions, which we account for during data preprocessing, especially for cognition-aware analysis.

\subsection{Preprocessing}

We processed the raw keystroke event sequences for each user into a structured DataFrame, as shown in Table \ref{tab:keystroke_sequence}. Each row represents a matched keydown–keyup pair, capturing \textit{question\_index}, \textit{key}, \textit{code}, keydown timestamp, keyup timestamp, and a continuity flag. The \textit{question\_index} follows an \textit{x.y} format, where \textit{x} denotes the typing scenario \textit{(1: bona fide, 2: LLM-response paraphrased, 3: LLM-response transcribed)} and \textit{y} indicates the specific question within that scenario.

We have introduced the continuity flag to identify the temporal discontinuity we noted earlier in the paper. The continuity flag identifies interruptions in the natural typing flow, particularly when a user revisits a previously answered question within the same typing scenario. This distinction is crucial for the accurate computation of Key Interval Time (KIT), which measures the time between consecutive key presses, particularly when keystroke data are grouped by \textit{question\_index} for cognitive load-aware analysis. While an upper threshold is applied to exclude excessively large KIT values resulting from such interruptions, the inclusion of the continuity flag provides further precision by explicitly identifying and excluding non-sequential transitions that could otherwise inflate KIT values.

\begin{table*}[htp]
\centering
\small
\renewcommand{\arraystretch}{1.15}
\begin{tabular}{p{2.2cm} p{1.7cm} p{1.7cm} p{2.3cm} p{2.3cm} p{1.7cm}}
\toprule
\textbf{question\_index} & \textbf{key} & \textbf{code} & \textbf{keydown} & \textbf{keyup} & \textbf{continuous} \\
\midrule
\texttt{$1.1$} & \texttt{ㅁ}         & KeyA       & 1723028762335 & 1723028762388 & TRUE \\
\texttt{$1.2$} & ㅏ & KeyK  & 1723028763839 & 1723028763840 & TRUE \\
\multicolumn{6}{c}{\textellipsis} \\
\texttt{$1.2$} & \texttt{ㄹ}         & KeyF       & 1723028853990 & 1723028854059 & TRUE \\
\texttt{$1.1$} & Backspace & Backspace  & 1723028854317 & 1723028854318 & FALSE \\
\bottomrule
\end{tabular}

\caption{Keystroke sequence illustrating temporal discontinuity assuming the user revisits question 1.1. When keystrokes are grouped by \textit{question\_index} $1.1$ without accounting for temporal discontinuities, the KIT between the \texttt{ㅁ} and Backspace keys would be $91,982$ ms, which could distort the resulting features.
}
\label{tab:keystroke_sequence}

\end{table*}
\vspace*{10pt}

\begin{figure*}
    \centering
    \includegraphics[width=7.5in, height=2.5in]{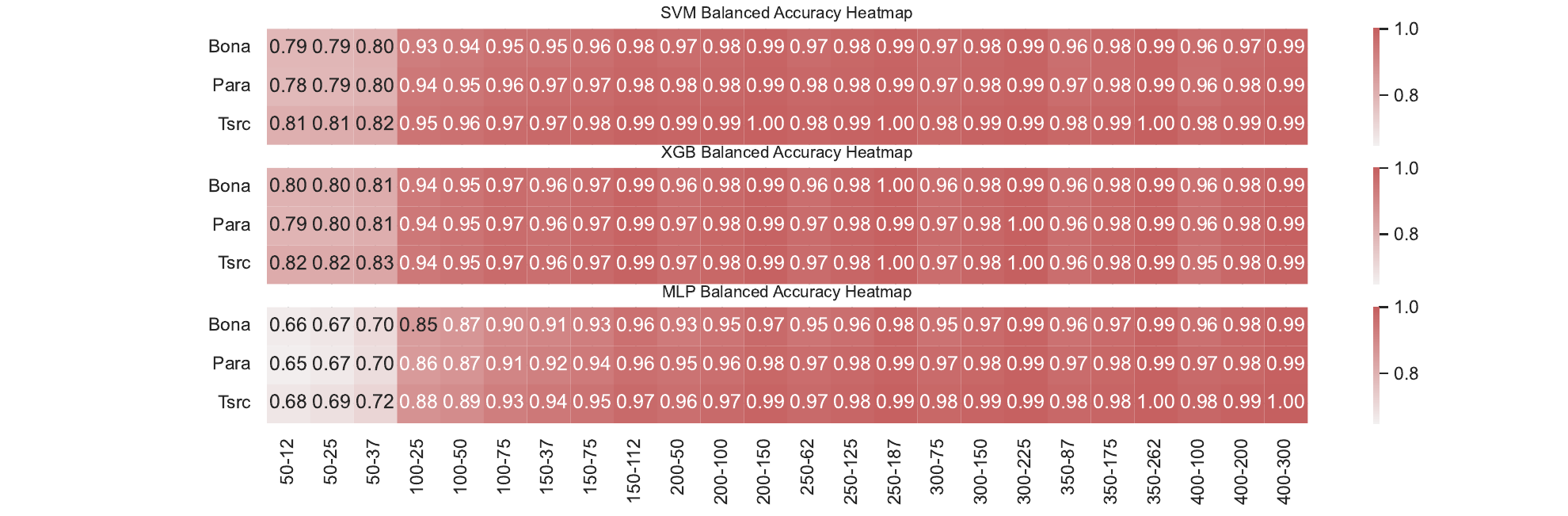}
    \caption{Heatmap of average balanced accuracy on training data across all users, evaluated over different combinations of window length and overlap. The results highlight the impact of segmentation parameters on model performance.}

    \label{fig:balanced-acc-wind-overlap}
\end{figure*}

\subsection{Feature extraction}

To extract meaningful temporal features from the keystroke sequences, we applied a sliding window approach to segment the data within each question. For each user, keystrokes were first grouped by \textit{question\_index} and sorted chronologically by keydown timestamp. The sequences were then divided into overlapping windows of fixed size, with window lengths ranging from 50 to 400 keystrokes (in increments of 50). For each window size, three overlap settings were used—25\%, 50\%, and 75\%—to preserve typing continuity across multiple scales. 


We extracted KIT and KHT features from each window for every combination of window size and overlap. KHT was calculated as the difference between the keyup and keydown timestamps for each key event. Because a key press and its corresponding release occur as a single uninterrupted action, this measure remains unaffected by temporal discontinuities, such as those introduced when a user revisits a previously answered question.

KIT was computed as the time interval between consecutive keydown events within each window. In contrast to KHT, KIT is sensitive to temporal discontinuities in the typing sequence (as illustrated in Table \ref{tab:keystroke_sequence}). To account for this, we implemented an algorithm to utilize the continuous flag assigned during preprocessing. When a continuous keydown event was followed by another continuous keydown, the interval between them was recorded as a valid KIT. If, however, the subsequent keydown was marked as non-continuous, the interval was excluded. In such cases, the algorithm advanced and computed KIT between the non-continuous event and the next available continuous keydown. This approach ensures that KIT is derived only from an uninterrupted input stream and avoids inflated intervals caused by participants revisiting previously answered questions.

\subsection{Feature matrix generation}
For each combination of window size and overlap, we generated training and testing sets using features extracted from Phase 1 and Phase 2 keystroke data, respectively. In doing so, we identified a set of common features—either single keys (for KHT) and key pairs (for KIT)—that had at least four valid values per window and appeared in at least 60\% of training windows across all users. This was done to preserve informative features while filtering out those with low occurrence, which could introduce noise.

For consistent feature dimension, we computed five summary statistics from each retained common feature: first quartile, median, third quartile, mean, and standard deviation, following \cite{roh2025llm}. Values were winsorized between the 10th and 90th percentiles to reduce the influence of outliers. Furthermore, missing values in training and testing sets were imputed using feature-wise means calculated from the training set.

\subsection{Window size and overlap selection}

To determine the optimal window size and overlap combination, we evaluated user authentication performance using stratified 5-fold cross-validation on the training data across all combinations. Balanced accuracy was used as the primary evaluation metric to account for imbalanced class distribution, and results were computed separately for each typing scenario. 

For each configuration and typing scenario, we modeled user authentication as a binary classification task, treating each individual as the genuine user (positive class) and all others as impostors (negative class).

Moreover, to assess classification performance, we employed a Support Vector Machine (SVM), a Multilayer Perceptron (MLP), and an Extreme Gradient Boosting (XGB) model. Both SVM and MLP pipelines included feature standardization, while XGB, being scale-invariant, was applied without standardization. The oversampling method SMOTE was applied only to MLP, as SVM and XGB can internally address class imbalance through mechanisms such as class weighting and loss adjustment. Additionally, hyperparameter tuning was conducted via grid search within each fold of a stratified 5-fold cross-validation.

As a result, balanced accuracy was computed per user and then averaged across users to obtain a single performance score for each configuration. Figure~\ref{fig:balanced-acc-wind-overlap} presents a heatmap of the average balanced accuracy computed on training data across all users for various combinations of window size and overlap. The results indicate that a window size of 200 with an overlap of 150 consistently yields the highest balanced accuracy across all three classifiers. This configuration offers an effective trade-off between performance and usability, as it maintains strong classification accuracy while requiring fewer keystrokes per segment—making it more practical for real-world applications.

\subsection{Feature selection}
Keystroke-derived features are high-dimensional and often redundant, especially when generated from overlapping windows and diverse user input \cite{FeatureSelectionKeystrokes}. To improve model generalization and reduce overfitting, we apply Minimum Redundancy Maximum Relevance (MRMR) with mutual information (MI) scoring. This approach selects features that are both individually informative and collectively diverse. Mutual information captures non-linear dependencies with the class label, while MRMR avoids selecting correlated features that add little discriminative value. Empirically, MRMR+MI outperformed univariate MI ranking and MRMR+(f-test) scoring on the training and validation data, yielding better cross-user and cross-context performance across all evaluation settings. The presented results, therefore, are based on features selected by the MRMR+MI method.  

\subsection{Classification algorithms} 
We selected three widely used machine learning classifiers—SVM, MLP, and XGB—based on their demonstrated effectiveness in keystroke-based behavioral biometrics~\cite{roh2025llm, Junhong2018Kdua, SVMKeystrokes, SVMKeystrokes2, SVM-1CKeystroke}. These models are well-suited for settings with moderate feature dimensionality and limited labeled data, which characterize the dataset used in this study. While recent deep learning approaches have shown promise in behavioral biometrics~\cite{acien2022, stragapede2024typeformer, Stragapede2023KVC, BehaveFormer}; their ability to generalize in small, heterogeneous settings remains limited ~\cite{Wahab2023, TreeBetterThanDLKeystrokes, MLBetterThanDeepKeystrokes}.

Each of our chosen classifiers addresses different aspects of the classification challenge. SVM is known for its robustness in high-dimensional, binary settings and is effective in situations with class imbalance \cite{SVMKeystrokes, SVM-1CKeystroke}. MLP offers non-linear modeling capacity with relatively low complexity \cite{MLBetterThanDeepKeystrokes}. XGB is particularly effective for tabular and sparse data, such as keystroke sequences, and has demonstrated consistent performance across behavioral tasks \cite{MLBetterThanDeepKeystrokes}. Together, these models provide a strong, interpretable baseline for evaluating context-aware authentication under constrained data settings.

\subsection{Evaluation scenarios}
We evaluate authentication performance across two independent factors: \textit{writing scenario} and \textit{cognitive load}. Each keystroke sample is labeled with one of three writing scenarios—\textit{bona fide typing}, \textit{LLM-response paraphrasing}, or \textit{LLM-response transcription}—and one of two cognitive load levels (\textit{Low} and \textit{High}). Those cognitive load levels are derived from Bloom's Taxonomy: \textit{Low} (remember, understand, apply) and \textit{High} (analyze, evaluate, create). 

The experiment was conducted under four setups: (1) scenario-unaware, cognition-unaware (SU-CU), (2) scenario-unaware, cognition-aware (SU-CA), (3) scenario-aware, cognition-unaware (SA-CU), and (4) scenario-aware, cognition-aware (SA-CA). For simplicity, we present the results in two parts: scenario-aware and scenario-unaware, with cognition-aware and cognition-unaware associated with each. 

\subsection{Evaluation metrics and curves}

We use Equal Error Rate (EER) as the primary evaluation metric. EER is the point at which the false acceptance rate (FAR) and false rejection rate (FRR) are equal. Lower EER indicates stronger authentication performance. To visualize classifier behavior across decision thresholds, we present Detection Error Tradeoff (DET) curves \cite{DETCombined} for all three typing scenarios—Bona fide, Paraphrased, and Transcribed—under various cognitive configurations. We also include Violin plots \cite{violinplot1996} to illustrate the distribution of EERs across individual users, capturing variability and robustness across classifiers and cognitive settings. These plots show that a few users have particularly high EERs, which is not uncommon in biometrics \cite{Doddington1998}. 
 
\begin{figure*}
    \centering
    \includegraphics[width=0.95\linewidth]{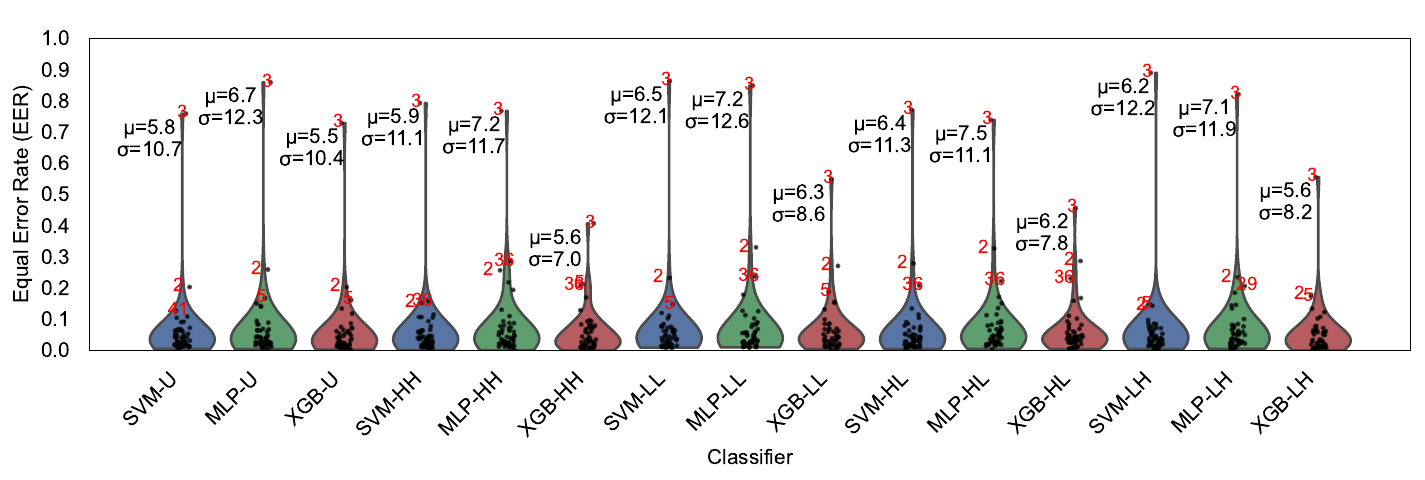}
    \caption{Distribution of EER(\%) across users for each classifier and cognitive context. Each violin shows the EER spread for individual users, with the mean and standard deviation annotated above. The three base classifiers (SVM, MLP, XGB) are colored consistently across all conditions. Outliers with the highest EER values are highlighted with user identifiers.}

    \label{fig:SU-EERs}
\end{figure*}

\begin{figure*}
    \centering
    \includegraphics[width=0.99\linewidth]{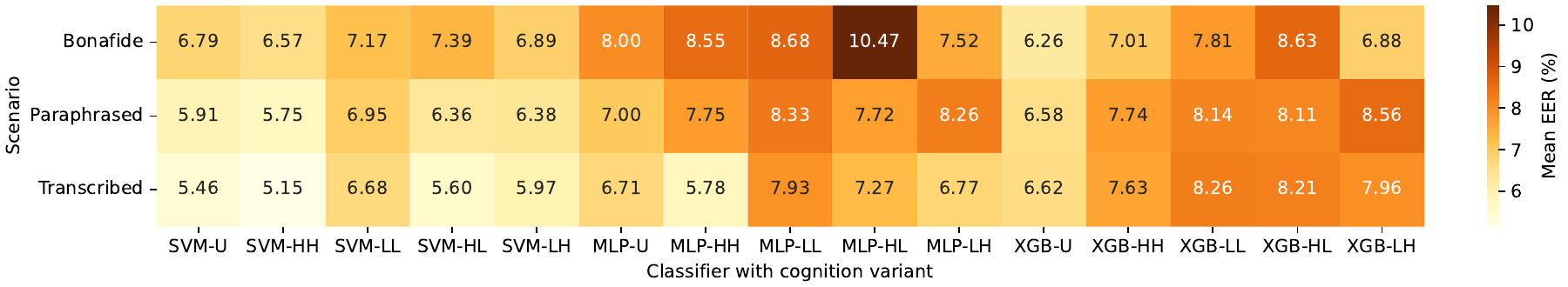}
    \caption{Mean EER(\%) across three typing scenarios for all classifiers and cognitive condition combinations. Variants are grouped by classifier family for better interpretation. Cognitive condition codes indicate training and test load levels:
U (Unaware): High+Low$\rightarrow$High+Low, HH: High$\rightarrow$High, LL: Low$\rightarrow$Low, HL: High$\rightarrow$Low, LH: Low$\rightarrow$High.
Lower values indicate better verification performance.}
    \label{fig:sa_eer}
\end{figure*}

\section{Results and discussions}
We present the results in two parts. The first part focuses on EER analysis across users, typing scenarios, cognitive load conditions, and classifier types, including both scenario-unaware (SU) and scenario-aware (SA) configurations. The second part presents operational performance using DET curves, with SVM selected as the designated classifier due to its consistent performance across conditions.

\subsection{Overall performance via EER}

\subsection*{ 1) Scenario-unaware configuration}

Figure~\ref{fig:SU-EERs} shows violin plots of user-wise EER distribution for each classifier and cognitive train–test setting. Each distribution includes 50 users, with mean ($\mu$) and standard deviation ($\sigma$) values annotated. The three classifiers—SVM, MLP, and XGB—are color-coded and evaluated under five cognitive configurations: U (Unaware: High+Low$\rightarrow$High+Low), HH (High$\rightarrow$High), LL (Low$\rightarrow$Low), HL (High$\rightarrow$Low), and LH (Low$\rightarrow$High), where $X$$\rightarrow$$Y$ denotes training on $X$ cognitive level and testing on $Y$ cognitive level.

\uline{\textit{SU classifier-level analysis}}
Across all conditions, XGB demonstrates the most consistent and reliable performance, achieving the lowest average EER ($\mu = 5.5\%$) in the U configuration and maintaining comparable scores in HH and LH (both $\mu = 5.6\%$). SVM closely follows, with strong results in U ($\mu = 5.8\%$), HH ($\mu = 5.9\%$), and LH ($\mu = 6.2\%$). In contrast, MLP underperforms in all settings, showing higher variability and mean EERs (e.g., $\mu = 6.7\%$ in U and $\mu = 7.5\%$ in HL). These findings highlight the superior generalizability of XGB and SVM in context-rich authentication, with XGB showing the tightest user distributions and lowest variance. 

\uline{\textit{SU cognition-level analysis}}
The HH configuration (SVM: 5.9\%, MLP: 7.2\%, XGB: 5.6\%) yields lower EERs than LL (SVM: 6.5\%, MLP: 7.2\%, XGB: 6.3\%), suggesting that higher cognitive load exhibits reduced intra-user and increased inter-user variability. Mismatched cognitive conditions—HL (SVM: 6.4\%, MLP: 7.5\%, XGB: 6.2\%) and LH (SVM: 6.2\%, MLP: 7.1\%, XGB: 5.6\%)—exhibit higher EERs than HH, but are comparable to LL in some cases. XGB handles mismatched conditions slightly better than MLP and SVM, which show larger degradation. The U condition, on the other hand, exhibits lower EERs than the cognition-aware (matched or mismatched) settings. However, this advantage may partly result from the increased amount of training and testing data available under U compared to cognition-aware settings, since U incorporates both high and low cognitive states, resulting in twice the amount of data. 

\uline{\textit{SU user-level analysis}}
A small subset of users, such as Users 2 and 3, consistently exhibit high EERs across classifiers and cognitive settings. These outliers, shown in Figure~\ref{fig:SU-EERs}, likely reflect typing behavior that is either highly variable or inherently more difficult to model. By contrast, the majority of users achieve EERs below 10\%, with most falling under 5\%, demonstrating the system’s reliability for typical users. In general, the lowest EERs are observed with XGB and SVM under the U and LH configurations.

In summary, XGB emerges as the most effective classifier in this part of the evaluation, with SVM a close second. Within the scenario-unaware context, cognitively diverse configurations deliver the best user-level generalization, underscoring the value of flexible, context-resilient training strategies in keystroke-based authentication.

\begin{figure*}[htp]
    \centering
    \begin{subfigure}[b]{0.24\textwidth}
        \includegraphics[width=\textwidth]{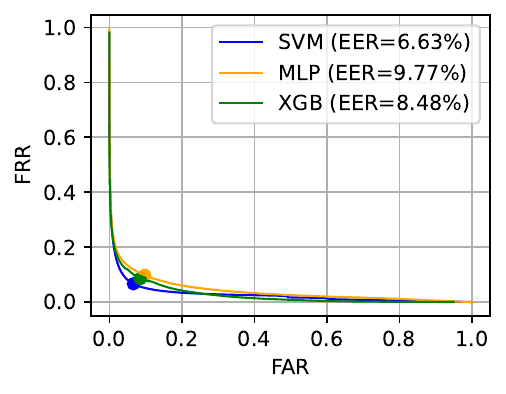}
        \caption{Scenario-Unaware}
        \label{fig:DET-SU-CLF}
    \end{subfigure}
    \begin{subfigure}[b]{0.24\textwidth}
        \includegraphics[width=\textwidth]{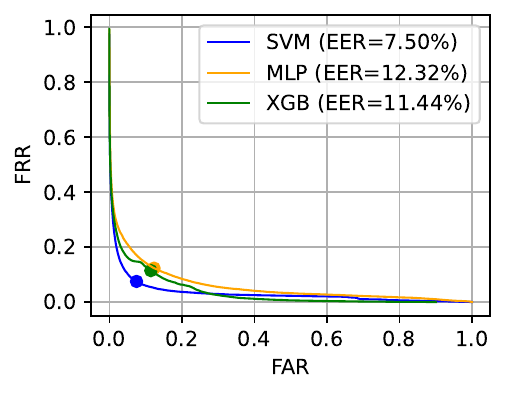}
        \caption{Scenario-Aware-Bona}
        \label{fig:DET-SA-CLF-B}
    \end{subfigure}
     \begin{subfigure}[b]{0.24\textwidth}
        \includegraphics[width=\textwidth]{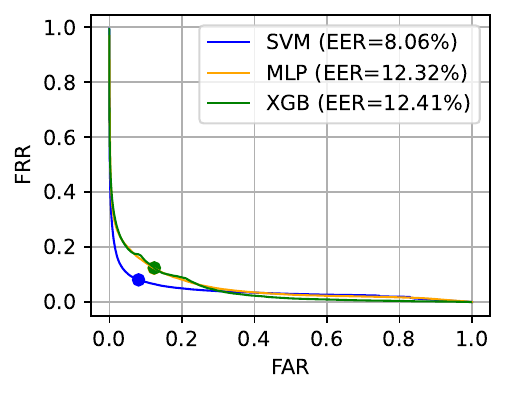}
        \caption{Scenario-Aware-Para}
        \label{fig:DET-SA-CLF-P}
    \end{subfigure}
        \begin{subfigure}[b]{0.24\textwidth}
        \includegraphics[width=\textwidth]{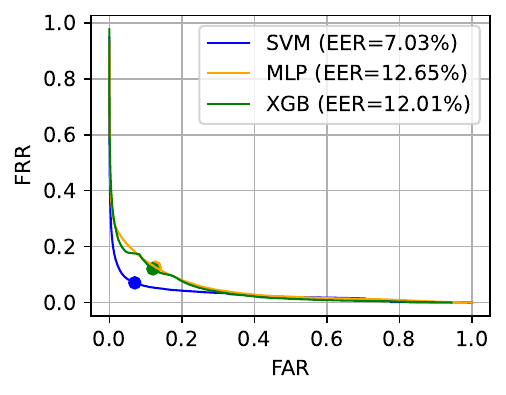}
        \caption{Scenario-Aware-Tsrc}
        \label{fig:DET-SA-CLF-T}
    \end{subfigure}
    \caption{Classifier-level DETs for Scenario-Unaware and Scenario-Aware Experiments.}
    \label{fig:Classifier-DETs}
\end{figure*}

\begin{figure*}[htp]
    \centering
    \begin{subfigure}[b]{0.24\textwidth}
        \includegraphics[width=\textwidth]{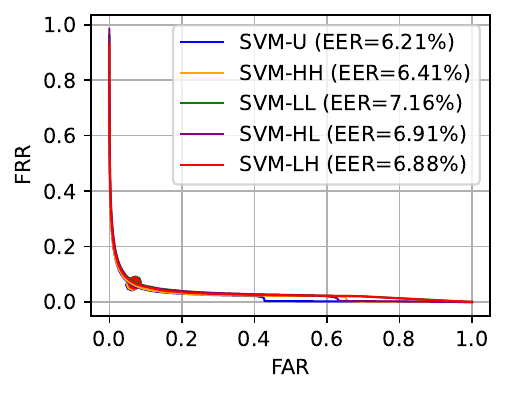}
        \caption{SVM-SU-Cognition}
        \label{fig:SVM-SU-Cognition}
    \end{subfigure}
    \begin{subfigure}[b]{0.24\textwidth}
        \includegraphics[width=\textwidth]{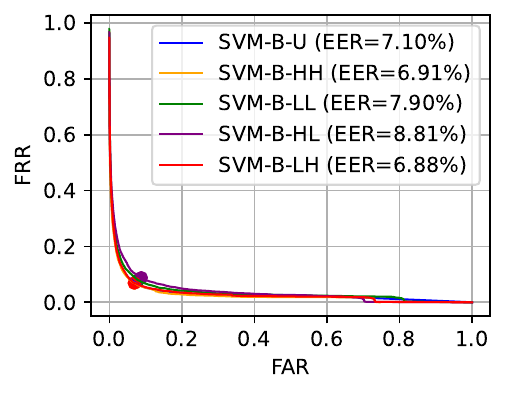}
        \caption{SVM-Bona-Cognition}
        \label{fig:SVM-Bona-Cognition}
    \end{subfigure}
    \begin{subfigure}[b]{0.24\textwidth}
        \includegraphics[width=\textwidth]{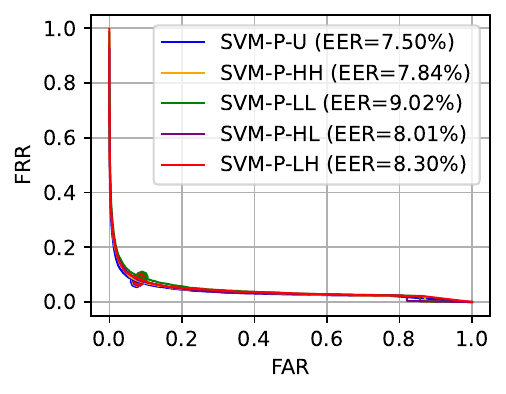}
        \caption{SVM-Para-Cognition}
        \label{fig:SVM-Para-Cognition}
    \end{subfigure}
    \begin{subfigure}[b]{0.24\textwidth}
        \includegraphics[width=\textwidth]{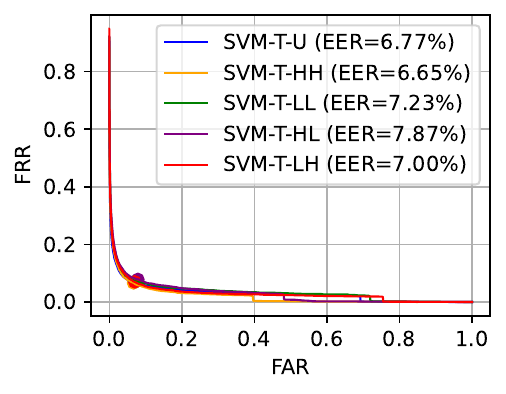}
        \caption{SVM-Tsrc-Cognition}
        \label{fig:SVM-Tsrc-Cognition}
    \end{subfigure}
    \caption{Cognition-level DETs for SVM in Scenario-Unaware and Scenario-Aware Experiments.}
    \label{fig:cognitionDETs}
\end{figure*}

\subsection*{ 2) Scenario-aware configuration}

Figure~\ref{fig:sa_eer} shows the mean EER across three typing scenarios—Bona fide, Paraphrased, and Transcribed—grouped by classifier (SVM, MLP, XGB) and cognitive configurations (U, HH, LL, HL, LH).

\uline{\textit{Scenario-level analysis}}  
Transcribed typing yields the lowest EERs across nearly all settings (e.g., SVM-HH: 5.15\%, MLP-HH: 5.78\%), suggesting that it produces more consistent typing behavior for each user, albeit with variations among users. Paraphrased performance lies between the other two conditions, possibly due to differences in the degree of paraphrasing across participants, which contributed to higher intra- and inter-user variance. Bona fide typing, being cognitively demanding and less constrained, shows the highest variability (e.g., MLP-HL: 10.47\%).

\uline{\textit{SA classifier-level analysis}}  
SVM is consistently strong, especially in Transcribed (SVM-HH: 5.15\%) and Paraphrased settings (SVM-U: 5.91\%). XGB is close behind with a strong Bona fide performance (XGB-U: 6.26\%) and low variance.  

\uline{\textit{SA cognition-level analysis}} Across typing modes, cognitive load has a consistent influence on authentication performance. In Bona fide inputs, EERs are the highest, with HH (SVM-HH: 6.57\%) and U (XGB-U: 6.26\%, SVM-U: 6.79\%) yielding the strongest results, while HL (MLP-HL: 10.47\%) produces the worst EER. LL and LH fall in an intermediate range but are less reliable across models. For Paraphrased inputs, SVM-HH (5.75\%) and SVM-U (5.91\%) again achieve the lowest errors, while HL and LH increase errors, with LH showing the steepest decline in XGB-LH (8.56\%) and MLP (8.26\%). Transcribed inputs are the easiest to classify. U and HH consistently produce strong results (SVM-U: 5.46\%, SVM-HH: 5.15\%), while LL leads to higher errors (SVM-LL: 6.68\%, MLP-LL: 7.93\%, XGB-LL: 8.26\%), although less extreme than in Bona fide. The cross-cognition performances (SVM-HL: 5.60\%, SVM-LH: 5.97\%) fall in an intermediate range.

\subsection{Operational performance via DET curves}
\subsection*{ 1) Classifier-level performance}

Figure~\ref{fig:Classifier-DETs} presents classifier-level DET curves for scenario-unaware and scenario-aware settings. We compare SVM, MLP, and XGB across four conditions: (a) Scenario-Unaware, (b) Scenario-Aware Bona fide, (c) Scenario-Aware Paraphrased, and (d) Scenario-Aware Transcribed.

\uline{\textit{Overall performance}}  
SVM consistently outperforms both MLP and XGB across all settings, achieving the lowest EER: 6.63\% in scenario-unaware, 7.50\% in Bona fide, 8.06\% in Paraphrased, and 7.03\% in Transcribed. MLP performs the worst, with EERs exceeding 12\% in all scenario-aware settings. XGB ranks in the middle but trails SVM by a clear margin (e.g., XGB: 12.41\% vs. SVM: 8.06\% in Paraphrased).

\uline{\textit{Scenario impact}}  
Scenario-unaware training leads to improved operational performance for all classifiers. For example, SVM achieves a reduction of approximately 1\% in EER compared to the best scenario-aware setting. This aligns with earlier findings that high cognitive load or mixed cognitive data improve generalization. Scenario-aware configurations—particularly Paraphrased—pose greater difficulty, likely due to increased variability in user behavior under paraphrased typing.

\uline{\textit{Classifier selection justification}}  
Given its consistent edge across all conditions, we select SVM as the default classifier for downstream deployment and cognition-aware analysis. Its DET curves remain steeper and lower across evaluation scenarios.

\subsection*{ 2) Cognition-level performance}

Figure~\ref{fig:cognitionDETs} shows cognition-specific DET curves for SVM across scenario-unaware and scenario-aware settings (Bona fide, Paraphrased, Transcribed). Each model reflects one of five cognitive training-testing configurations: U (Unaware), HH, LL, HL, and LH. 

In the \uline{\textit{Scenario-unaware}} case, U (EER = 6.21\%) slightly outperforms HH ($6.41\%$) and LH (6.88\%), with LL performing worst (7.16\%). This suggests that training on mixed cognitive data (U) and data collected under high-cognitive load (HH) supports better generalization for authentication models. In contrast, low-load-only training may underrepresent behavioral variability. 

In the \uline{\textit{Bona fide}} scenario, LH (6.88\%) and HH (6.91\%) outperform HL (8.81\%) and LL (7.90\%). Models trained on low-load and tested on high-load (LH) benefit from the stability of low-load input, while the HL drop hints at a mismatch cost when training on noisier, high-load keystrokes. In the \uline{\textit{Paraphrased}} case, U (7.50\%) and HH (7.84\%) are strongest, while LL (9.02\%) degrades sharply. This aligns with the expectation that paraphrased text introduces lexical and syntactic complexity, requiring exposure to higher cognitive diversity during training. For the \uline{\textit{Transcribed}} scenario, HH (6.65\%) and U (6.77\%) outperform LH (7.00\%) and HL (7.87\%), indicating that transcription benefits from high-load and mixed cognitive conditions. 

Overall, high-load (HH) or mixed cognitive conditions (U) yield robust generalization. In contrast, keystroke patterns collected under low-cognitive load exhibit high authentication errors, likely due to increased inter-user similarity. These trends affirm the importance of both cognitive variability and directionality in model generalization for keystroke-based verification.
 
\section{Conclusion and Future Work}
This work establishes the feasibility of keystroke-based active authentication in Korean under realistic writing conditions shaped by varying LLM usage and cognitive demands. By systematically evaluating performance across bona fide, paraphrased, and transcribed input, as well as across cognitive levels derived from Bloom’s Taxonomy, we demonstrate that cognitively diverse datasets or data collected under high cognitive load help the model generalize more effectively. In contrast, data collected under low-cognitive load exhibit a high error rate. Among the evaluated classifiers, SVM demonstrates consistent superiority across conditions, making it a strong candidate for deployment in context-aware behavioral authentication.

Looking forward, this study opens several directions for extending keystroke-based authentication. Richer writing domains, including multilingual, conversational, and creative contexts, can provide new behavioral signals and test model generalizability. Incorporating adaptive or online learning strategies may further strengthen long-term resilience to intra-user variability. Finally, combining keystroke dynamics with complementary behavioral signals, or leveraging LLM detection as part of the authentication process, presents promising opportunities for building context-aware and cognitively sensitive security systems.

\balance
\bibliography{references}

\end{document}